\documentclass[twocolumn,showpacs,aps,prb,superscriptaddress,floatfix]{revtex4}

\usepackage{graphicx}
\begin{document}
%
%
\author{S. R. Dunsiger}
\affiliation{Department of Physics and Astronomy, McMaster University, Hamilton, Ontario, L8S 4M1, Canada.}
\author{Y. Zhao}
\affiliation{Department of Physics and Astronomy, McMaster University, Hamilton, Ontario, L8S 4M1, Canada.}
\author{Z. Yamani}
\affiliation{Canadian Neutron Beam Centre, NRC, Chalk River Laboratories, Chalk River, Ontario, K0J 1J0, Canada}
\author{W. J. L. Buyers}
\affiliation{Canadian Neutron Beam Centre, NRC, Chalk River Laboratories, Chalk River, Ontario, K0J 1J0, Canada}
\affiliation{Canadian Institute for Advanced Research, 180 Dundas St. W.,
Toronto, Ontario, M5G 1Z8, Canada}
\author{H. Dabkowska}
\affiliation{Department of Physics and Astronomy, McMaster University, Hamilton, Ontario, L8S 4M1, Canada.}
\author{B. D. Gaulin}
\affiliation{Department of Physics and Astronomy, McMaster University, Hamilton, Ontario, L8S 4M1, Canada.}
\affiliation{Canadian Institute for Advanced Research, 180 Dundas St. W.,
Toronto, Ontario, M5G 1Z8, Canada}
\title{Incommensurate Spin Ordering and Fluctuations in underdoped La$_{2-x}$Ba$_{x}$CuO$_{4}$}
\date{\today }
\newcommand{\LBCO}{La$_{2-x}$Ba$_x$CuO$_{4}$}
\newcommand{\LSCO}{La$_{2-x}$Sr$_x$CuO$_{4}$}
\newcommand{\LBCOy}{La$_{2-x}$Ba$_x$CuO$_{4+y}$}
\newcommand{\LSCOy}{La$_{2-x}$Sr$_x$CuO$_{4+y}$}
\newcommand{\tc}{T$_C$}
\begin{abstract}
Using neutron scattering techniques, we have studied incommensurate spin ordering as well as 
low energy spin dynamics in single crystal underdoped \LBCO\ 
with x$\sim$0.095 and 0.08; high temperature superconductors with T$_C \sim$ 27 K and 29 K respectively.
Static two dimensional incommensurate magnetic 
order appears below T$_N$=39.5 $\pm$ 0.3 K in \LBCO (x=0.095) and a similar temperature for x=0.08 within the 
low temperature tetragonal phase.  
The spin order is unaffected by either the onset 
of superconductivity or the application of
magnetic fields of up to 7 Tesla applied along the c-axis in the x=0.095 sample. 
Such magnetic field {\it independent} behaviour
is in marked contrast with the field induced enhancement of the staggered magnetisation observed in the 
related \LSCO\ system, indicating this phenomenon is not a universal property of cuprate superconductors.
Surprisingly, we find that incommensurability $\delta $ is only weakly dependent on doping relative to \LSCO.
Dispersive excitations in \LBCO\ (x=0.095) at the same 
incommensurate wavevector persist up to at least 60 K.   The dynamical spin susceptibility 
of the low energy spin excitations saturates below \tc, in a similar manner to that seen in the superconducting 
state of La$_2$CuO$_{4+y}$.
\end{abstract}
\maketitle
%
%
The roles of spin, charge and lattice degrees of freedom have been central to the rich behaviour 
brought to light over the last 20 years in superconducting lamellar copper
oxides~\cite{kastner98,tranquadareview,birgeneau06}.   
In particular, the cuprates exhibit phenomena which are a sensitive function of doping, 
evolving from an antiferromagnetic insulating parent compound into a superconducting phase with 
increasing hole density.  
A heterogeneous electronic phase composed of stripes of 
 itinerant charges now appears to be a generic feature of hole
 doped ternary transition metal oxides~\cite{neto03} such as manganites~\cite{manganites1,manganites2} and 
 nickelates~\cite{nickelates1,nickelates2,nickelates3}, as well as cuprates.
The explanation for these incommensurate spin ordered states is the subject of ongoing debate.
In an itinerant picture, the spin dynamics are described in terms of electron-hole pair
excitations about an underlying Fermi surface~\cite{bulut90,bulut96,norman00}.
Alternatively, within the stripe picture of doped, two dimensional Mott insulators, the non-magnetic holes in these 
materials organize into quasi-one dimensional stripes which separate antiferromagnetic insulating antiphase 
domains~\cite{kivelson03}.  Adjacent antiferromagnetic regions are $\pi$ out of phase with each other 
giving rise to a magnetic structure with incommensurate periodicity, where the supercell dimension is twice the hole stripe periodicity.

The static spin structures in the undoped, parent compounds such as 
La$_2$CuO$_4$~\cite{vaknin87} or YBa$_2$Cu$_3$O$_6$~\cite{tranquada88} have been determined by neutron scattering 
to be relatively simple two sublattice antiferromagnets
characterized by a commensurate ordering wavevector of (0.5,0.5) 
in reciprocal lattice units within the tetragonal basal plane.
On hole doping with either Sr substituting for La in La$_{2-x}$Sr$_x$CuO$_4$~\cite{incommensurate1} 
or by adding additional oxygen in 
YBa$_2$Cu$_3$O$_{6+x}$~\cite{stock04,stock06}, the magnetic scattering moves out to incommensurate wavevectors, 
consistent with the 
stripe ordering picture described above.  This incommensurate magnetism can be 
either static or dynamic, as evidenced by either elastic or inelastic peaks in the neutron scattering 
respectively and now appears to be a common feature of the \LSCO\  
family of compounds.  
Specifically, for lightly doped La$_{2-x}$Sr$_x$CuO$_4$, elastic incommensurate magnetic Bragg features first 
appear split off
from the (0.5,0.5) position in diagonal directions relative to a tetragonal unit cell~\cite{matsuda00,wakimoto00}.  
At higher doping in the underdoped superconducting regime, the peaks 
rotate by 45$^{\circ }$ to lie along directions parallel to the tetragonal axes or 
Cu-O-Cu bonds, such that
elastic magnetic scattering appears at (0.5$\pm$$\delta$,0.5,0) and (0.5,0.5$\pm$$\delta$,0)~\cite{suzuki98,kimura99}. 
For optimal and 
higher doping the static order disappears, but dynamic incommensurate correlations 
nevertheless persist\cite{kimura99,wakimoto04}.   

Within the stripe picture, one expects charge 
ordering associated with the holes to occur at an incommensurate wavevector $2\delta $, twice that describing the spin order. 
Neutron scattering is not directly sensitive to charge ordering {\it per se}, but it is sensitive to atomic 
displacements, such as those associated with oxygen, which arise from charge ordering.  
An incommensurate nuclear scattering signature is therefore expected to 
appear at (2$\pm$2$\delta$,0,0) or (2,$\pm 2\delta$,0) and 
related wavevectors. Despite extensive efforts, such incommensurate charge related scattering has not been observed in 
La$_{2-x}$Sr$_x$CuO$_4$ by either neutron or X-ray scattering techniques, although there is
indirect evidence of charge stripe excitations from optical measurements~\cite{lucarelli03}.
Such scattering {\it has} been observed in La$_{1.6-x}$Nd$_{0.4}$Sr$_x$CuO$_4$~\cite{tranquada95,tranquada97}, as 
well as La$_{1.875}$Ba$_{0.125-x}$Sr$_x$CuO$_4$  (x=0,0.05,0.06,0.075,0.085)~\cite{fujita02},
motivating discussion as to whether such static charge stripes compete with, rather than
underlie, high temperature superconductivity.  
\begin{figure}
\centering
\includegraphics[angle=90,width=0.9\columnwidth,clip]{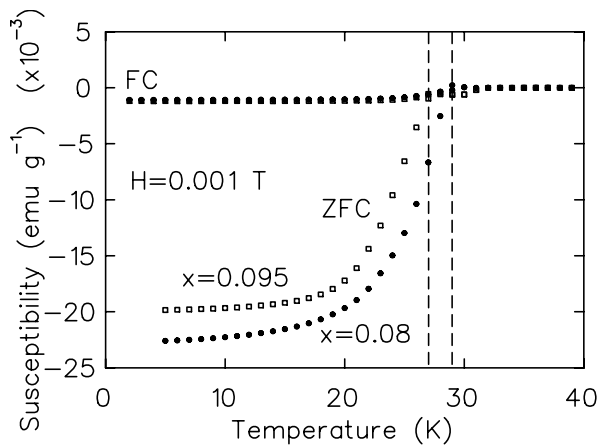}
\caption{Zero field cooled (ZFC) and field cooled (FC) susceptibilities of \LBCO (x=0.095 and 0.08) single crystals
measured at 0.001 T.  Dashed lines indicate the onset of superconductivity at
T$_C$=27 and 29 K for x=0.095 and x=0.08 samples respectively.} 
\end{figure}
\begin{figure}
\centering
\includegraphics[angle=0,width=0.85\columnwidth,clip]{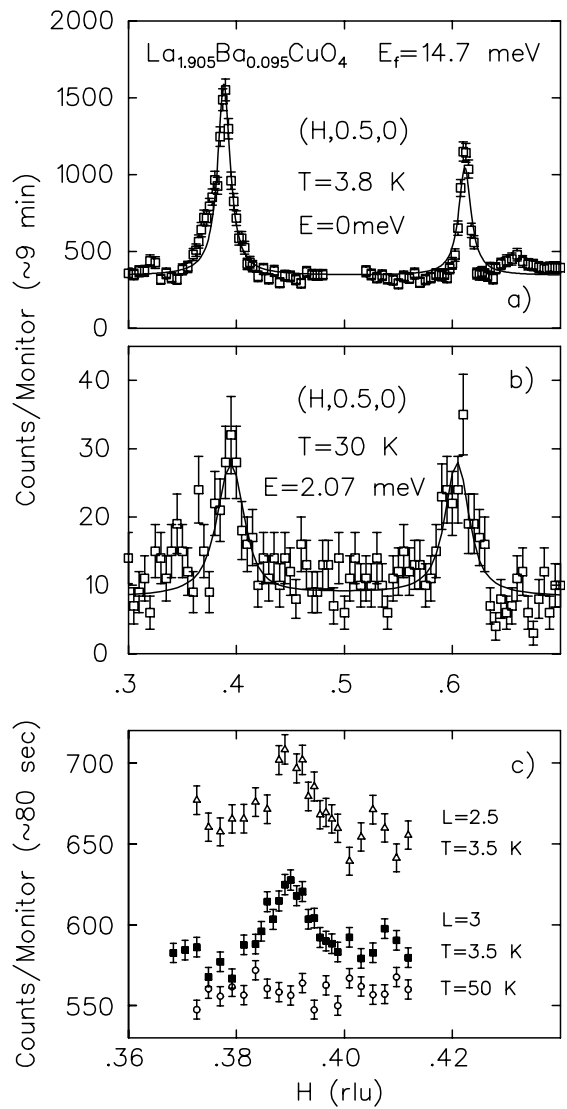}
\caption{a) Static incommensurate magnetic peaks with $\delta$=0.112 in La$_{1.905}$Ba$_{0.095}$CuO$_4$ at T=3.8 K, 
along (H,0.5,0).  b) Representative inelastic scans at T=30 K, also along (H, 0.5, 0) and at $\hbar\omega$=
2.07 meV.  Parameters characterizing this inelastic scattering are shown in Fig. 4, while the solid line
is discussed in the text.  c) 
Elastic scans of the form (H, H/(1-2$\delta $), L), which demonstrate the rod-like, two dimensional 
nature of the elastic magnetic scattering, as described in the text.  The scan at L=2.5 has been displaced by 50 counts upwards for
clarity.} 
\end{figure}
Surprisingly, La$_{2-x}$Ba$_x$CuO$_4$,  the first high temperature superconductor (HTSC) to be 
discovered~\cite{bednorz86}, has been 
much less extensively studied than either \LSCO\ or YBa$_2$Cu$_3$O$_{6+x}$
due to the difficulty of growing single crystals, which has only been achieved\cite{adachi01} recently. 
In this paper, we report neutron scattering signatures of static incommensurate spin order in single crystal 
\LBCO\ (x=0.095, 0.08), consistent with the stripe picture described above, but with 
interesting complexity not accounted for within present theoretical models.  Our results clearly indicate 
that the spin ordering is insensitive to both the onset of superconductivity and surprisingly the application of a magnetic
field.
 Tranquada {\it et al}~\cite{Tranquada04} and Fujita {\it et al}~\cite{fujita04}
have reported neutron scattering measurements on an x=0.125 sample of La$_{2-x}$Ba$_x$CuO$_4$.  In 
both \LBCO\ and \LSCO\ this 
concentration corresponds to a suppression of T$_C$ as a function of doping,      
known as the ``1/8" anomaly.  In \LBCO\ the suppression is almost 
complete~\cite{moodenbaugh88} and is associated 
with a
structural phase transition at low temperature, from orthorhombic 
to tetragonal~\cite{axe89} which gives rise to a superlattice peak at 
(0,1,0) and symmetry related reflections.  
Samples of \LBCO\ near x=0.125 display a sequence of structures on 
lowering the 
temperature, going progressively from high temperature tetragonal (HTT, I4/mmm symmetry) to orthorhombic (MTO, bmab symmetry) to 
low temperature tetragonal (LTT, P4$_2$/ncm symmetry)~\cite{axe89}. 
The high temperature tetragonal to orthorhombic transition in particular, and to a lesser extent, the
orthorhombic to low temperature tetragonal transition are sensitive indicators for the precise Ba doping level in the material.  

\section{Single Crystal Growth, Characterization and Experimental Procedure}
We have grown~\cite{Yang} high quality single crystals of \LBCO\ with x=0.095 and x=0.08
using floating zone image furnace techniques with a four-mirror optical furnace.  A small single 
crystal of La$_2$CuO$_4$ was employed as a seed for the growth, which was performed under enclosed pressures 
of 165 and 182 kPa of O$_2$ gas for x=0.095 and x=0.08 samples respectively.  
The resulting $\sim 6$ gram single crystals of \LBCO\ were cylindrical in
shape and cut to dimensions 25 mm in length by 5 mm in diameter (x=0.095) and 38 mm in length by 5 mm diameter (x=0.08). 
We have determined the Ba concentration from the HTT to MTO transition temperature.
The \LBCO (x=0.095) and (x=0.08) single crystals displayed HTT to MTO structural phase transitions at T$_{d1}$$\sim$272 K and
T$_{d1}$$\sim$305 K, respectively, and MTO to LTT transitions at T$_{d2}$$\sim$45 K 
and T$_{d2}$$\sim$ 35 K, respectively~\cite{Yang}.  The bulk superconducting transition temperatures T$_C$=27 K (x=0.095) and T$_C$=29 K (x=0.08)
are identified using the onset of the zero field cooled diamagnetic response of the crystals, as measured using 
Superconducting Quantum Interference Device (SQUID) magnetometry and shown in Fig. 1.  The various structural and 
superconducting phase transition temperatures are summarised in Table 1, along with the corresponding values for the 
La$_{2-x}$Ba$_x$CuO$_4$ (x=0.125) sample~\cite{fujita04}.

Oxygen stoichiometry in \LBCOy\ is more difficult to quantify, especially at the low levels relevant here.  
Experience with \LSCOy\ suggests the oxygen stoichiometry, y, is negative in as-grown samples, giving rise to crystals 
which possess an effective doping level which is lower than that given by the Sr concentration alone; the effective doping 
level is $x+2y$ \cite{yamada98}.  Stoichiometric samples at optimal and underdoped Sr concentrations display a maximum superconducting T$_C$ 
which cannot be increased by controlled annealing in O$_2$ gas.  Excess oxygen can be incorporated into LaCuO$_{4+y}$ crystals, 
but only through 
electrochemical doping methods, in which case y can be as high as 0.11 and the interstitial oxygen organizes itself into
staged structures~\cite{lee02}.    
\begin{table}
\centering
\caption{\label{tab:table1} Summary of \LBCO\ structural (T$_{d1}$, T$_{d2}$), superconducting (T$_C$) and magnetic (T$_N$)
phase transition temperatures~\cite{Yang}}
\begin{ruledtabular}
\begin{tabular}{ccccc}
x&$T_{d1}$(K)&$T_{d2}$(K)&$T_{c}$(K) & T$_N$(K)\\
\hline
0.125 & 232 & 60 & $\sim$4\footnotemark[1] & 50 \\
0.095 & 272& 45 & 27 & 39.5 \\
0.08& 305 & 35 & 29 & ~39 \\
\end{tabular}
\end{ruledtabular}
\footnotetext[1]{From ~\cite{fujita04}.}
\end{table} 

Neutron scattering experiments were undertaken on the C5 and N5 triple axis spectrometers at the 
Canadian Neutron Beam Centre, Chalk River.  All experiments were performed with pyrolytic graphite 
(002) planes as monochromator and analyser, with constant E$_f$ =14.7 meV.  A graphite filter was placed in the
scattered beam to reduce contamination from higher order neutrons.  The single crystals were
oriented with (H,K,0) in the horizontal scattering plane.  Crystallographic  indices are denoted using
tetragonal notation, where the basal plane lattice constant a=3.78 \AA\ at low temperatures.

\section{Elastic Neutron Scattering in L\lowercase{a}$_{2-x}$B\lowercase{a}$_x$C\lowercase{u}O$_{4}$ (\lowercase{x}=0.095)}
\subsection{Meissner State}
\begin{figure}[h]
\centering
\includegraphics[angle=0,width=0.9\columnwidth,clip]{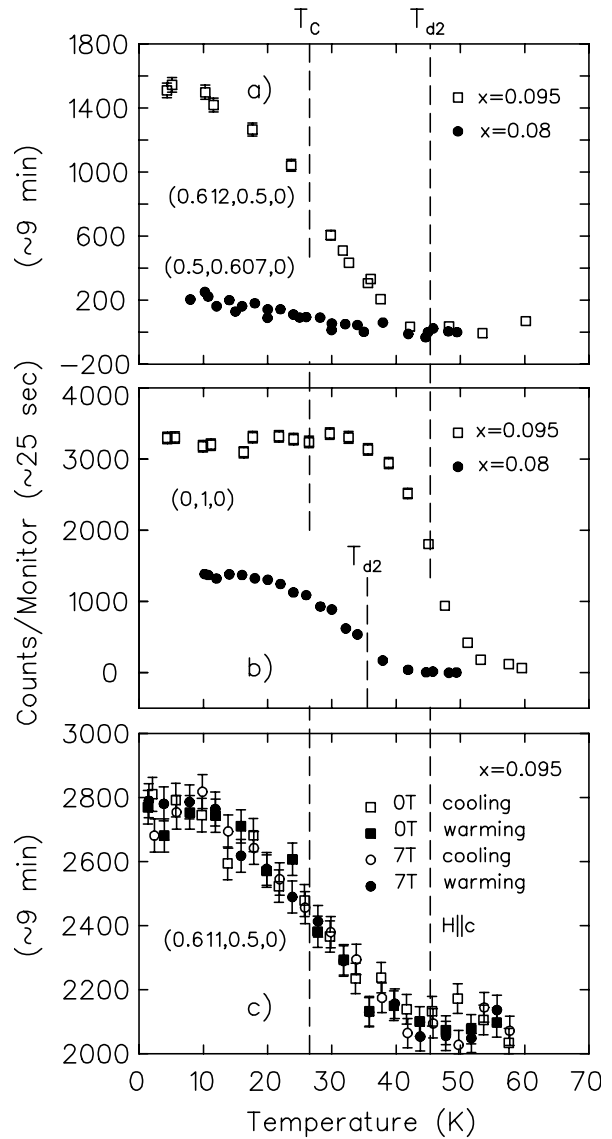}
\caption{a) The temperature dependence of the net elastic incommensurate magnetic scattering in \LBCO\ (x=0.095)
at (0.612,0.5,0) and (x=0.08) at (0.5,0.607,0), as well as that of
b) the (0,1,0) structural Bragg peak, which
marks the orthorhombic to low temperature tetragonal structural phase transition.
Note that a constant background has been subtracted in both cases.
The superconducting and structural
phase transition temperatures are indicated by dashed lines for both samples.
  All the data for x=0.08 have been scaled to the volume of the x=0.095 sample by phonon normalization.
 The counting times refer to the unscaled x=0.095 data.
 c) Temperature dependence of the elastic incommensurate magnetic scattering in \LBCO\ (x=0.095) in 0
and H=7 T $\parallel $ c.}
\end{figure}

We discuss the more extensive measurements on the \LBCO\ sample with x=0.095 first.
Elastic scattering scans at T=3.8 K are shown in Fig 2a), in which Bragg peaks occur at (0.5$\pm \delta $,0.5,0) 
$\delta =0.112(3) $,
indicating static incommensurate spin order.
Analogous magnetic Bragg peaks are observed  at (0.5,0.5$\pm$0.112,0).  
All peak widths are
resolution limited with full widths at half maximum, FWHM=0.011 \AA$ ^{-1}$, indicating static spin 
correlations within the basal plane
exceeding 180 \AA.  
By contrast, the spin correlations between planes are very short. 
To observe this, the crystal was reoriented in the (H,H,L)
scattering plane and then tilted $\sim 7^{\circ }$ at constant L to intersect the incommensurate peak 
position for H=0.39.  
Measurements along H of the form 
(H,H/(1-2$\delta $),L) at fixed L for L=2.5 and 3 are shown in Fig. 2c at 
T=3.5 K and at L=3 for T=50 K to extract a background.  
Since the peak intensity is independent of L, the scattering taking the form of an elastic 
rod along the L direction, the static spin order at low temperatures is two dimensional.
Note that the intensity in Fig. 2a) is greater relative to that in Fig. 2c) for the sample oriented in the (H,K,0) plane.  
This arises because the neutron spectrometer has a broad vertical resolution which integrates the signal 
in the L direction, perpendicular to the scattering plane.

The temperature dependence of the incommensurate magnetic elastic scattering is 
illustrated in Fig. 3a.  
The intensity of the magnetic Bragg peak is proportional to the volume average of the square of the 
ordered staggered moment.
The spin order at (0.612,0.5,0) develops continuously with temperature below
T$_{N}$=39.5$\pm$0.3 K.  The temperature dependence of the (0,1,0) structural Bragg peak indicates 
the orthorhombic to low temperature tetragonal phase transition occurs at T$_{d2}$$\sim$45 K, a 
transition which is 
discontinuous in nature~\cite{axe89}.  For reference, the superconducting transition 
at T$_C$$\sim$ 27 K (see Fig. 1) 
is also indicated on this plot as a dashed line.  The onset of spin ordering, T$_N$ correlates most strongly with the 
completion of the transition to the low temperature tetragonal phase, and the incommensurate spin order 
coexists with the superconductivity below T$_C$.  
Associated incommensurate charge ordering has not been observed. 
The temperature dependence of the spin ordering is qualitatively similar to that observed in the x=1/8
compound~\cite{fujita04}, where the superlattice peak intensity becomes non-zero below $\sim 50 $ K.  
Similarly, no anomaly has been observed at \tc\ in YBCO$_{6.35}$ in the spin order, 
which has been attributed to robust spin correlations~\cite{stock06}.

\subsection{Magnetic Field Dependence}
The most surprising result of this study is that the incommensurate spin structure shows no magnetic field dependence 
up to 7 T, applied vertically along the c* axis.  
 Neither cooling nor warming the sample in a magnetic field has an effect on either the 
 temperature dependence of the spin ordering, or the Bragg intensity in \LBCO\ (x=0.095), as shown in Fig. 3c.  
 This result is in 
 marked contrast with the behaviour of underdoped and optimally doped \LSCO\ where pronounced field dependent 
 effects are observed.  
For the optimally doped \LSCO\ compound (x=0.163), the application of a 
magnetic field enhances the dynamical spin susceptibility but does not induce 
static order~\cite{lake01}.  
Most dramatically, in a slightly underdoped sample (x=0.144),   
Khaykovich {\it et al}~\cite{khaykovich05} report the development of  a static incommensurate spin structure 
above a critical field of  2.7 T.  The authors therefore argue that \LSCO\ (x=0.144) may be tuned 
through a quantum critical point,
at which there is a magnetic field induced transition between magnetically disordered and ordered phases.
Their results are interpreted in terms of a Ginzburg-Landau model due to Demler {\it et al}~\cite{demler01}, which
assumes a microscopic competition between spin and superconducting order parameters.  
The predicted magnetic intensity increases as $\Delta I \sim H/H_{c2} \ln (H_{c2}/H)$~\cite{demler01}, which is
consistent with experiments on \LSCO.  Note that the intensity changes most rapidly with magnetic field at 
low fields on a scale set by $H_{c2}$. 
In the \LBCO\ family of compounds, the lower critical field is H$_{c1}\sim 0.04$ T, 
while the upper critical field H$_{c2}$ is in excess of 40 T~\cite{takagi87}.  The upper critical field is of the same order of 
magnitude in optimally doped \LSCO~\cite{boebinger96}.  Thus an applied magnetic field of 7 T should be sufficiently large 
to see an effect in \LBCO.   

For sufficiently underdoped \LSCO\ (x=0.12,0.10), 
the ordered magnetic moment associated with preexisting static spin order is enhanced on application 
of a magnetic field~\cite{katano00,lake02,lake05}. The spin order within the vortex state 
of \LSCO\ (x=0.10)~\cite{lake02} indicates long in-plane
correlation lengths, greater than both the superconducting coherence length
and the intervortex spacing  at 14.5 T.  As the coherence length is a measure of the size of the vortices, Lake {\it
et al} argue the static magnetism must therefore reside beyond the extent of the vortices themselves~\cite{lake02}.  
Whereas the \LBCO\ (x-0.095) correlation length for static spin order is 
similarly long, the underlying physics is clearly
different and the spins appear to order independent of vortex creation.

\section{Inelastic Neutron Scattering in L\lowercase{a}$_{2-x}$B\lowercase{a}$_x$C\lowercase{u}O$_{4}$ (\lowercase{x}=0.095)}
\begin{figure}[h]
\centering
\includegraphics[angle=0,width=0.9\columnwidth,clip]{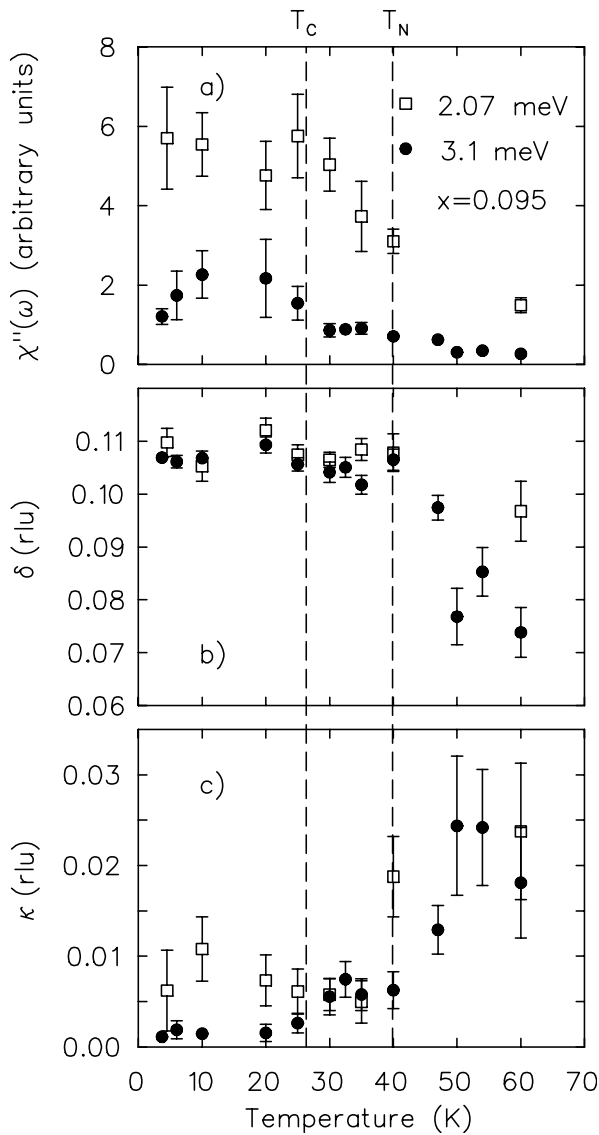}
\caption{The temperature dependence of the parameters extracted from fitting the low energy inelastic 
magnetic scattering shown in the middle panel of Fig. 2.  This scattering was fit to Eq. 1, and we show
(top panel) $\chi "$({\bf Q}, $\hbar\omega$=2.07 and 3.1 meV); (middle panel) the incommensurability 
$\delta$; and (lower panel) the inverse correlation length $\kappa$.  The dashed lines indicate the 
superconducting (T$_C\approx 27$ K) and magnetic (T$_N$=39.5 $\pm$0.3 K)
transition temperatures.}
\end{figure}
The magnetic excitations were studied in constant energy 
transfer scans performed through 
the incommensurate ordering wavevectors.  Horizontal collimation 
sequences of 0.54$^{\circ }$-0.48$^{\circ }$-S-0.54$^{\circ }$-1.2$^{\circ 
}$ 
and 0.54$^{\circ }$-0.79$^{\circ }$-S-0.85$^{\circ }$-2.4$^{\circ }$ were used at
energy transfers of 2.07 and 3.1 meV respectively, yielding corresponding energy resolutions of $\sim 1$ and
$\sim 1.5$ meV FWHM.  The representative scan
along (H,0.5,0) and $\hbar\omega$=2.07 meV at T=30 K in Fig. 2b, shows that the low energy 
dynamic spin response peaks up at the same wavevector, (0.5$\pm$0.112,0.5,0), as the static spin structure. 
At higher energy transfers the signal declines rapidly.   
The measured dynamic structure factor $S({\mathbf Q},\omega )$  is related to the imaginary part of 
the dynamical susceptibility $\chi '' ({\mathbf Q},\omega )$  
through the fluctuation-dissipation theorem.
For quantitative analysis, the data have been fit to the resolution convolution of 
$S({\mathbf Q},\omega )= \chi '' ({\mathbf Q},\omega )[1-\rm{e}^{-\hbar \omega /k_B T}]^{-1}$
where the susceptibility~\cite{fujita04} is:
\begin{equation}
\chi '' ({\mathbf Q},\hbar\omega )=\chi '' (\hbar\omega ) \sum _{n=1}^4
\frac{\kappa}{({\mathbf Q}-{\mathbf Q}_{\delta, n})^2 +\kappa ^2} 
\end{equation}
and ${\mathbf Q}_{\delta, n}$ represents the four incommensurate wave vectors 
$(\frac{1}{2}\pm \delta ,\frac{1}{2},0)$ and $(\frac{1}{2},\frac{1}{2}\pm \delta ,0)$.
This assumes the magnetic excitations consist of four rods of scattering running along the c$^*$ axis.
The extracted temperature dependences of $\chi ''(\hbar\omega)$, 
$\delta$ and $\kappa$ are plotted in Figs. 4a, b and c respectively.  $\chi '' (\hbar\omega ) $ is proportional to 
the integral of 
$\chi '' ({\mathbf Q},\omega )$ over ${\mathbf Q }$ in the (H,K,0) scattering plane; $\delta$ is the incommensurability,
while $\kappa$ is the inverse of the static correlation length in the basal plane, defined as the peak half width at 
half maximum.  For reference, both the spin ordering transition at T$_N$$\sim$ 39.5$\pm$0.3 K, and the superconducting 
transition near T$_C$$\sim$ 27 K are indicated on this plot.   
At both 2.07 meV and 3.1 meV, the dynamical susceptibility, $\chi '' (\hbar\omega ) $,  increases
continuously as the 
temperature is reduced below $\sim$ 60 K, becoming roughly constant and non-zero below T$_C$ $\sim$ 27  K.  
 This is similar 
to measurements in both overdoped La$_2$CuO$_{4+y}$, where a levelling off of the dynamic incommensurate spin response 
has been reported below T$_C$$\sim$ 42 K~\cite{lee99} and also in \LBCO\ (x=0.125) in the normal
state~\cite{fujita04}.  In the latter compound, as a function of frequency there is relatively little change in  
$\chi '' (\hbar\omega ) $ at low temperature (8 K), whereas it drops rapidly in the present x=0.095 sample.  
As the temperature is raised, $\chi '' (\hbar\omega ) $  
varies linearly with frequency at lower energy transfers below 10 meV in the x=0.125 sample for T$>65$ K, 
whereas it declines with increasing $\omega $ in x=0.095 for all
T$<60$ K.  Note that the signal intensity has been corrected for the monitor sensitivity to higher order incident neutrons, as described
in Ref.~\cite{stock04}.
These low energy excitations have 
some characteristics of the spin waves observed in the parent compound La$_2$CuO$_4$~\cite{yamada89} as one warms through
the N{\'e}el temperature, where instantaneous spin correlations with the character of the N{\'e}el state persist into
the paramagnetic regime~\cite{shirane87}.

The form of $\chi '' (\hbar\omega ) $ varies dramatically as a function of doping in the related \LSCO\
compounds.  In optimally and slightly overdoped \LSCO (x=0.15,0.18)~\cite{yamada95,yamada97} there is a 
characteristic energy of $\sim 7$ meV below which the dynamic susceptibility is dramatically reduced in the 
superconducting state - the opening up of a spin gap.  However, on the underdoped side of the superconducting dome there 
is finite spectral weight in the spin response at all low energy transfers~\cite{lee00,hiraka01}.

The bottom two panels of Fig 4 show that the incommensurability, $\delta$ and the inverse 
correlation length, $\kappa$, correlate most strongly with the disappearance of the static spin order 
near T$_N$$\sim$39 K, which is not surprising. 
Above T$_N$ the increase in $\kappa $ may indicate that stripe correlations are weakened by thermal fluctuations that broaden
the hole distribution about antiphase domain boundaries.

\section{Elastic Neutron Scattering in L\lowercase{a}$_{2-x}$B\lowercase{a}$_x$C\lowercase{u}O$_{4}$ (x=0.08)} 
%
%
\begin{figure}[h]
\centering
\includegraphics[angle=90,width=0.9\columnwidth,clip]{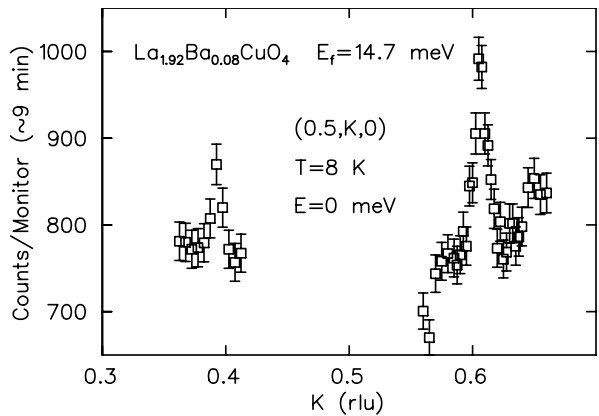}
\caption{Static incommensurate magnetic peaks with $\delta$=0.107 in La$_{1.92}$Ba$_{0.08}$CuO$_4$ at T=8 K, 
along (0.5,K,0).}
\end{figure}
Qualitatively, the magnetic and superconducting properties of \LBCO\ (x=0.08) (T$_C$=29 K) are very similar
to those of the higher doped x=0.095 (T$_C$=27 K) sample.   
 Elastic neutron scattering measurements were carried
out under the same conditions as described earlier, with the single crystal oriented with (H,K,O) in the
horizontal scattering plane.
The elastic scattering scans at T=8 K (see Fig. 5) show that the incommensurate wave vector has decreased from 
$\delta = 0.112 (3)$ in the x=0.095 sample to 0.107(3). 
Surprisingly, the magnetic scattering is roughly a factor of eight less intense (Fig. 3a).  
In Figs. 3a) and b) the intensities have been scaled to the sample volume using the
integrated intensity of an acoustic phonon measured near a strong nuclear Bragg peak at (2,0.15,0).  
We do not understand the reduced intensity for x=0.08, since extinction does not play a role.

The temperature dependence of the elastic magnetic signals at the incommensurate wavevectors (0.5,0.607,0)
for \LBCO\ (x=0.08) and at (0.612,0.5,0) for \LBCO\ (x=0.095) are reproduced in Fig. 6, where
the intensities have been scaled so that their functional form may be directly compared.  As can be seen, both 
the temperature dependence of the order parameter and the phase transition temperatures are
very similar despite the difference in the strength of the elastic magnetic Bragg scattering.
We therefore conclude the two electronic energy scales for these crystals at x=0.08 and x=0.095, set by the 
superconducting T$_C$ and 
T$_N$, are surprisingly similar.  It is not clear why a $4\% $ decrease in $\delta $ should produce an 
eight-fold decrease in
the spin Bragg intensity.  A search revealed no additional magnetic intensity in diagonal directions.
As in x=0.095, no incommensurate peaks due to charge ordering were observed in the x=0.08 sample. 

%
%
\begin{figure}[h]
\centering
\includegraphics[angle=90,width=0.9\columnwidth,clip]{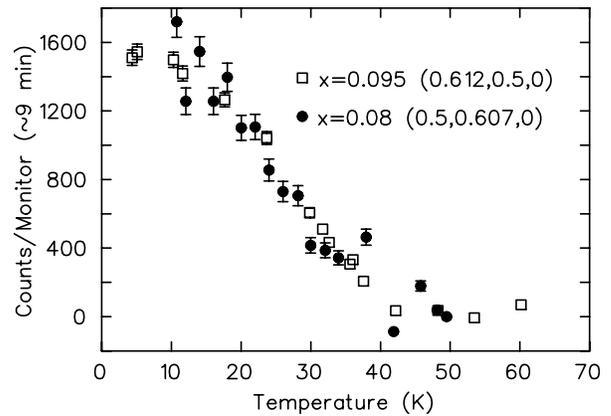}
\caption{The temperature dependence of the elastic incommensurate magnetic scattering in \LBCO\ (x=0.095) 
at (0.612,0.5,0) and (x=0.08) at (0.5,0.607,0).  The net intensity for x=0.08 has been multiplied by a factor of  7.7
to match that of x=0.095.}
\end{figure}
\section{Discussion}
Whether magnetism and superconductivity coexist in the same 
microscopic regions of the CuO$_2$ planes or are phase separated is a topical subject of research.
The issue of microscopic spatial segregation has been examined 
using a combination of neutron scattering~\cite{khaykovich02} and $\mu $SR~\cite{savici02} techniques in  LaCuO$_{4+y}$.
 As a local probe $\mu $SR is sensitive to heterogeneous structures.
 The magnetic ordering in LaCuO$_{4+y}$ (y=0.11) and  \LSCO\ (x=0.12) has been reported to occur in  reduced magnetic volume 
fractions of 40 and 18$\%$ respectively~\cite{savici02}.  Khaykovich {\it et al} argue an applied magnetic field enhances
spin ordering primarily in the nonmagnetic regions~\cite{khaykovich02}, consistent with the above observations.
By contrast, the magnetic volume fraction in \LBCO\ (x=0.095) is much larger, approaching 100$\% $~\cite{dunsiger07}.  
We speculate that no
magnetic field dependence has been observed in \LBCO\ x=0.095 because the non-magnetic volume fraction is too low.
A systematic study of the variation of the spin ordering with magnetic field is therefore of interest, with 
emphasis on the correlations between this effect  and the magnetic volume fraction. 

An interesting difference between the x=0.08 and the 0.095 \LBCO\ samples
is that the spin ordered state in the x=0.095 sample grows within a fully
developed LTT structure, as T$_N$$\sim$39.5 K and T$_{d2}$$\sim$ 45 K.  
By contrast, in the x=0.08 \LBCO\ sample, the MTO to LTT structural phase
transition begins near T$_N$ on decreasing temperature and is only
completed at temperatures below $\sim$ 20 K.  The situation for x=0.125
\LBCO\ is similar to the x=0.095 case, as T$_N$$\sim$ 50 K, at which
temperature the MTO-LTT transition for x=0.125 is largely complete.  The
first order nature of the MTO-LTT structural phase transition implies
co-existing structures over the temperature regime at which the spin order
forms for x=0.08 \LBCO.  It is then possible that the resulting structural
heterogeneity interferes with the full development of spin order, giving
rise to a substantially reduced magnetic Bragg intensity as compared with
the x=0.095 sample.  However, we also note that variability\cite{lake02} in the elastic
magnetic Bragg intensity has been reported from \LSCO\ sample to sample
with similar nominal doping levels of x$\sim$ 0.1 and the \LSCO\ system does not
display the LTT phase at low temperatures.
\begin{figure}[h]
\centering
\includegraphics[angle=0,width=0.9\columnwidth,clip]{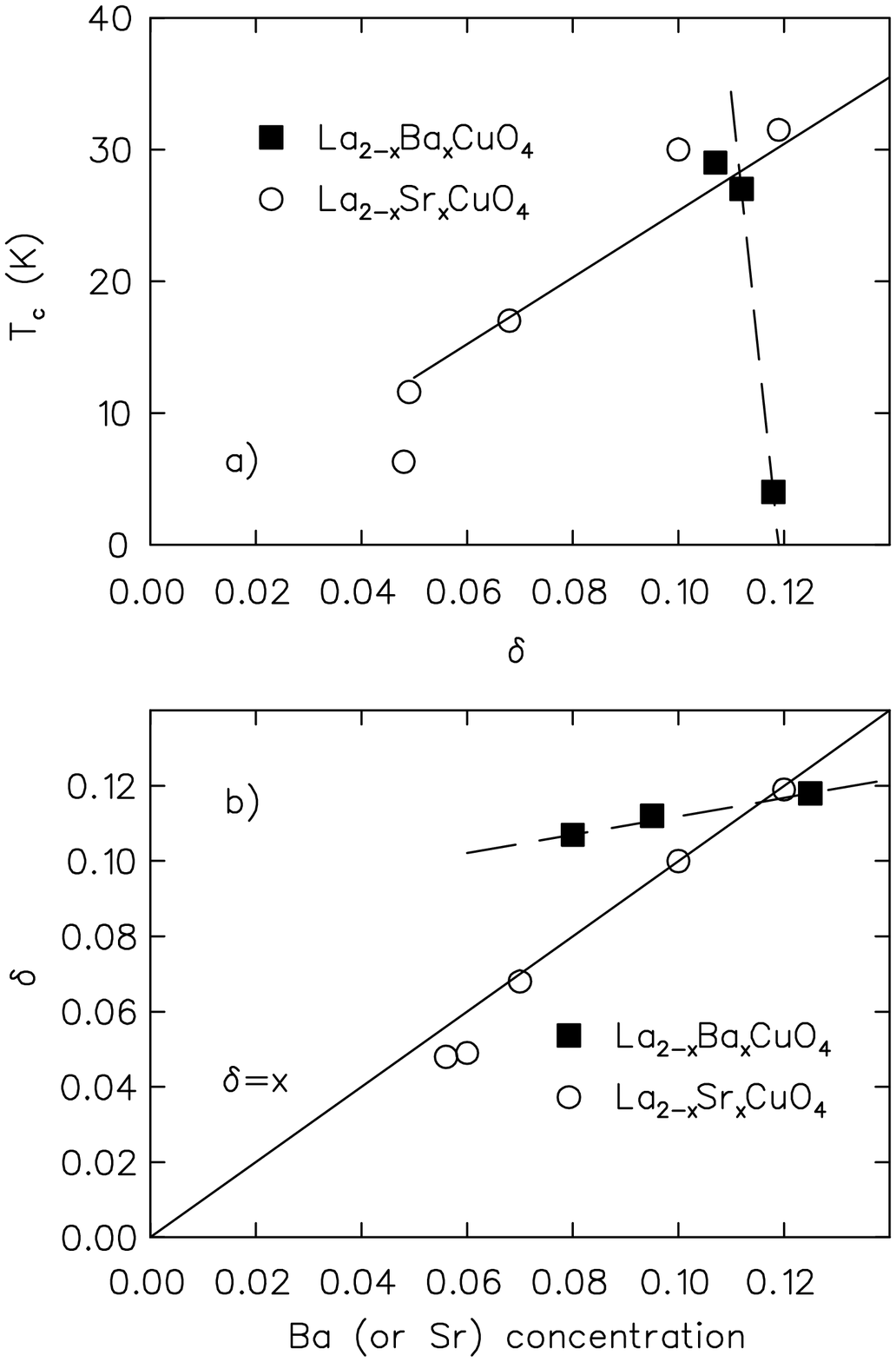}
\caption{a) Superconducting \tc\ is plotted vs elastic incommensuration, $\delta$, for the 
x=0.08, 0.095, and 0.125 \LBCO\ samples, compared with those measured in related \LSCO\ samples.
b)  Incommensurability $\delta$ is plotted vs Ba-content, x, for our x=0.08, x=0.095 \LBCO\ samples,
with previous x=0.125 \LBCO\ results~\cite{fujita04} and those relevant to several underdoped \LSCO\ samples~\cite{kimura99, fujita02a}.  The
dotted line is a guide to the eye.}
\end{figure}

It is also interesting to examine the correlation between Ba-content, x, incommensurability, $\delta$, and superconducting \tc\ 
in \LBCO\ and compare these relationships to those reported for \LSCO.  The top panel of Fig. 7 shows 
\tc\ vs $\delta$ for the x=0.08, 0.095, and 0.125 \LBCO\ samples, compared with those measured in related \LSCO\ samples.  
As can be seen, our results for the x=0.08 and x=0.095 samples give the same \tc\ as \LSCO\ for the same 
incommensuration, $\delta$. Of course, \LBCO\ displays the pronounced x=1/8 anomaly and consequently the x=0.125 point on this 
\tc\ vs $\delta$ plot lies far below the linear curve which is an excellent descriptor for the remainder of the 
underdoped \LBCO\ and \LSCO\ systems.  The abrupt nature of the x=1/8 anomaly in the LBCO system is clear.

The bottom panel of Fig. 7 shows the incommensurability $\delta$ vs Ba-content, x and once again we compare our new results on
x=0.08 and x=0.095 \LBCO\ samples with the x=0.125 \LBCO\ results and those of several underdoped \LSCO\ samples at Sr concentrations 
above x=0.05, where the incommensurate spin ordering is consistent with a picture of
parallel (as opposed to diagonal) stripe ordering.  In this Sr-content regime, $\delta$ tracks x well, assuming stoichiometric 
oxygen content~\cite{yamada98,kimura99,matsuda00,wakimoto00}.   One can see that the incommensuration 
in x=0.08 and x=0.095 \LBCO\ shows relatively little x-dependence. 
Indeed, we observe $\delta$ values which are only $\sim$ 9 $\%$ less than that displayed by x=0.125 \LBCO, thereby departing 
significantly from the approximately linear $\delta$ vs x relation characterizing the underdoped \LSCO\ studies.  

It is possible that this difference between underdoped \LBCO\ and \LSCO\ also arises due to the MTO-LTT structural phase transition, occurring 
in \LBCO\ but absent in \LSCO.  It is also conceivable that it arises due to some small oxygen off-stoichiometry, such that our 
samples have the composition \LBCOy, with oxygen stoichiometry greater than 4.  Such excess oxygen would give rise to an 
effective hole doping given by x$_{eff}$ = x+2y.  To bring the $\delta$ values for x=0.08 and 0.095 back onto the
linear relationship between $\delta$ and x$_{eff}$ seen in \LSCO, small, but positive values of y: 0.013 and 0.0075 for the 
x=0.08 and x=0.095 \LBCOy\ samples respectively would be required.  This is too small to be detectable and runs counter to what is 
concluded in \LSCOy.  In underdoped \LSCOy, the 
superconducting \tc\ is maximized by annealing in oxygen, at which point the measured $\delta$ vs Sr concentration, x, 
lie on the straight line \cite{yamada98}.  Consequently, as grown \LSCOy\ tends to be oxygen deficient (y$<$0) and annealing in 
oxygen results in 
stoichiometric \LSCO.  This is also expected to be true for underdoped \LBCOy, which would imply that the deviation of 
$\delta$ vs x from a linear relationship is intrinsic to stoichiometric \LBCO, a surprising result.

\section{Summary and Conclusions}
We have observed the coexistence of static, two dimensional incommensurate spin order and 
superconductivity in La$_{2-x}$Ba$_x$CuO$_4$ with x=0.095 and x=0.08.
This result is in broad agreement with other well studied high temperature
La-214 cuprate superconductors, such as La$_{2-x}$Sr$_{x}$CuO$_4$ or La$_2$CuO$_{4+y}$, 
which show signatures of either incommensurate static spin ordering or dynamic spin correlations.  One significant finding of this study 
is the field {\it independence} of the incommensurate magnetic order in the x=0.095 sample, 
in marked contrast with other superconducting  La-214 cuprates.  Studies of the spin ordering as a function of magnetic field in other 
superconducting
systems with large magnetic volume fractions should prove illuminating.
In addition, while the dependence of the superconducting \tc\ on the incommensuration of the 
magnetic structure, $\delta$, in \LBCO (x=0.08, 0.095) is the same as that observed in \LSCO, the x-dependence appears 
to be substantially weaker than that seen in \LSCO, where a linear relationship is observed over this
range of concentration. 
Now that crystal growth breakthroughs have resulted in the availability of large, high quality single crystals
of \LBCO, fuller experimental characterization of the original family of high temperature
superconductors is clearly warranted.  

This work was supported by NSERC of Canada.  It is a pleasure to acknowledge the contributions of A. Kallin, A. Dabkowski
and E. Mazurek at McMaster, who were involved in the single crystal growth and characterization and the
expert technical support of L. McEwan and R. Sammon at NRC, Chalk River.

\end{document}